**C-axis phonons in Fe-As Based Superconductors Investigated by Inelastic X-ray Scattering**

D. Reznik[1], K. Lokshin[2], D. C. Mitchell[3], D. Parshall[3], W. Dmowski[2], D. Lamago,[1,4] R. Heid[1], K.-P- Bohnen[1], A. S. Sefat[5], M. A. McGuire[5], B. C. Sales[5], D. G. Mandrus[5], A. Asubedi[3,5], D. J. Singh[5], A. Alatas[6], M. H. Upton[6], A. H. Said[6], Yu. Shvyd'ko[6], and T. Egami[2,3,5]

[1] Forschungszentrum Karlsruhe, Institut für Festkörperphysik, D-76021 Karlsruhe, Germany

[2] Department of Materials Science and Engineering, University of Tennessee, Knoxville, TN 37996

[3] Department of Physics and Astronomy, University of Tennessee, Knoxville, TN 37996

[4] Laboratoire Leon Brillouin, CE Saclay, F-91191 Gif-sur-Yvette, France

[5] Oak Ridge National Laboratory, Oak Ridge, TN 37831

[6] Advanced Photon Source, Argonne National Laboratory, Argonne, IL 60192

**In Fe-As based superconductors magnetism and superconductivity show strong sensitivity to the lattice, suggesting a possibility of unconventional electron-phonon coupling. We investigated *c*-axis polarized phonons by inelastic X-ray scattering. Their peak positions and linewidths did not vary significantly as a function of doping either by Co or by K. The linewidth of the Fe vibration shows unusual wavevector-dependence, which may be due to hybridization with in-plane modes. Comparison with the density functional theory shows significant difference for the energy of the As-As vibrations (Raman active at the zone center) but not for the Fe-As vibrations (infrared-active at the zone center). This behavior cannot be explained by a 10% softening of Fe-As interaction strength as proposed previously for in-plane polarized vibrations.**

PACS #:

The recent discovery of superconductivity in iron pnictides [1] with the critical temperature, $T_C$, up to 55 K [2] created huge excitement, since the superconductivity in these compounds is clearly unconventional, and, just as in the cuprate superconductors, cannot be explained by the standard BCS theory. They are poor metals [3], and magnetic and structural phase transitions are found in compositions in close proximity [4-7]. After the initial discovery [1] superconductivity was observed in a large number of closely related compounds. They all share the FeAs layers, in which Fe is surrounded by four As ions forming a tetrahedron, and $FeAs_4$ tetrahedra are connected by sharing the faces. The FeAs layers are separated by ionic layers, such as $REO_{1-x}F_x$ (RE = rare earths) or $Ba_{1-y}K_y$. The principal role of the separation layer appears to be maintaining the two-dimensional nature of the FeAs layer and doping charge carriers into it. However, as demonstrated by the size effect of the rare earth elements on $T_C$ [2], the separation layer also affects the FeAs layer through the lattice structure. Density functional theory (DFT) calculations obtain a very weak conventional electron-phonon coupling [8], which is not sufficient to explain the observed high values of $T_C$.

However the DFT calculations as well as experiments show remarkable sensitivity of superconductivity and magnetism to the lattice. Changing the ionic size of the rare-earth without changing the apparent charge density, $T_C$ changes by a factor of two [2], and the appearance of magnetism is closely tied to very small changes in the lattice parameter [3,9]. In addition, there is a large overestimate of the magnetic moments in generalized gradient approximation (GGA) calculations, an overestimate that appears essential for obtaining a correct As height in structure optimizations [10]. This necessity of including magnetism to obtain reasonable structures persists into the non-magnetic superconducting phase. Furthermore, the results of core level spectroscopy show exchange splitting even in the non-magnetic phases [11], similar to the $Fe_2Nb$, which is near an itinerant magnetic quantum critical point [12]. It may also be expected that magntoelastic couplings lead to modification of the magnetic phase diagrams. For example, they often convert soft second order transitions into first order transitions.

Sorting out the interplay between the lattice and magnetic degrees of freedom is a key challenge for understanding these materials and perhaps for understanding the superconductivity. For this purpose we used inelastic x-ray scattering to measure frequencies and linewidths of the phonons with atomic vibrations along the *c*-axis in double-layered (122) FeAs compounds, $BaFe_2As_2$, $Ba_{1-x}K_xFe_2As_2$ ($x = 0.2, 0.4$), and $BaFe_{1.8}Co_{0.2}As_2$, and compared the results to DFT calculations. We show that some phonon frequencies are in excellent agreement with the calculated ones whereas others are 20% lower in experiment than in the calculations.

The inelastic x-ray scattering (IXS) measurements were carried out at the Advanced Photon Source (APS), Argonne National Laboratory. The main insights were obtained based on the first set of measurements on $Ba_{0.8}K_{0.2}Fe_2As_2$ at the XOR 30-ID (HERIX) beamline and were confirmed by carrying out more detailed measurements with a smaller statistical error on the other three samples on the XOR 3-ID beamline. [13] The incident energy was about 23.72 keV (30-ID) or 21.66 (3-ID), and the horizontally scattered beam was analyzed by a horizontal array of nine/four spherically curved silicon crystals on 30-ID/3-ID, respectively. The energy resolution was 1.5/2.2 meV full width at half maximum (FWHM). The focused beam size of about 30/100 *μm* on 30-ID/3-ID enabled studying phonons even in a small crystal. Here we focus on the results for $BaFe_2As_2$ (FeAs), $Ba_{0.6}K_{0.4}Fe_2As_2$ (BaK), and $BaFe_{1.8}Co_{0.2}As_2$ (FeCo). The FeAs crystal was nonsuperconducting, whereas the doped samples had $T_c$s of around 30 K. In the preparation of crystals, high purity elements (> 99.9 %) were used and the source of all elements was Alfa Aesar. BaK crystals were grown out of a Sn flux. The procedure involved placing elements in the ratio of $[Ba_{0.6}K_{0.4}Fe_2As_2]$:Sn = 5:95, in an alumina crucible, and sealing into a fused silica tube under partial argon atmosphere (~1/3 atm). The sealed ampoule was placed in a furnace and heated at 800 C for 6 hours, then cooled (10 °C/hr) to 525 °C at which point the Sn was decanted off by means of a centrifuge. All the crystals had smooth surfaces with typical dimensions of ~ 2 x 3 x 0.1 $mm^3$. Single crystals of FeAs and FeCo were grown out of FeAs flux [14]. For FeAs, crystals prepared with a ratio of Ba:FeAs = 1:5 were heated at 1180°C under the conditions described above, for 8 hours. For FeCo, crystals with a ratio of Ba:FeAs:CoAs = 1:4.45:0.55 were similarly heated to 1180 °C, and dwelled for 10 hours. Both ampoules were cooled at the rate of 3 °C/hour, followed by decanting of FeAs flux at 1090 °C. The typical crystal sizes from both batches were ~ 5 x 5 x 0.2 $mm^3$. All crystals were well formed plates with the [001] direction perpendicular to the plane of the crystals. The crystals were placed in a closed cycle He refrigerator in the reflection geometry with the *(110)*- and *(001)*-axes in the horizontal plane to observe phonons with the polarization vector along the *c*-axis.

Figure 1 shows the spectra of BaFe$_2$As$_2$ (FeAs) at $T$ = 5 K, at wavevectors $Q$ = ($H$ = 0.04, $K$ = 0.04, $L$ = 4.87), (0, 0, 5.53), (0.04 0.04 6.2), and (0.07, 0.07, 6.87) in the units of reciprocal lattice vertors r.l.u. They were measured simultaneously using 4-analyzers/detectors with each analyzer corresponding to a different wavevector. The $H$, $K$ values are very close to the zone center whereas the $L$-values correspond to different parts of the zone along the $c^*$. In the range of the measured $L$-values $L$ = 6 is at the zone center and $L$ = 5 and 7 are at the zone boundary. Our phonon peak assignments are based on comparison with DFT calculations and previous Raman scattering results. Several phonon branches are seen in addition to the elastic line centered at zero energy. The lowest-energy branch disperses upward from low energies towards 8meV at the zone boundary and can be assigned to the acoustic branch. The next highest branch near 12meV shows a small downward dispersion and can be assigned to the optic branch due to the vibrations of the heaviest element, Ba. The next highest branch is the As-As out-of-phase vibration, which is Raman active at the zone center. It has a small downward dispersion (22meV at $L$ = 6.2) (20meV at $L$ = 4.87 and $L$ = 6.87). The highest energy branch (~ 35 meV) is the Fe-As out-of-phase vibration, which is IR-active at the zone center, is nearly dispersionless. Approximate eigenvectors of the two highest-energy branches are shown in the bottom row of Fig. 1. However, comparison of their measured intensities to the prediction of structure factor calculations shows that the eigenvectors shown in Fig. 1 ignore some Ba contribution, which is probably $q$-dependent. To perform a more accurate measurement of the eigenvectors more Brillouin zones must be measured. We note that the branch originating from the Fe Raman-active vibration of $B_{1g}$ symmetry has zero structure factor at the wavevectors where the measurements were performed. Since the interesting physics is believed to occur in the Fe-As layer we focused the rest of our study on the two As-As and Fe-As phonons, which correspond to the vibrations of the atoms in the Fe-As layer.

Figure 2 compares low temperature spectra of the parent non-superconducting compound (FeAs) and the two superconducting samples (BaK and FeCo). The 23 meV As-As peak is composition-independent, whereas the 35 meV Fe-As mode is broader and harder in the BaK sample than in the other samples. We also find the As-As mode is nearly temperature-independent in all samples (Data not shown). The Fe-As mode, however, may have a temperature dependent shift as 0.5 meV as illustrated in Fig. 3. The data are suggestive but not definitive on this point, due to possible systematic error. This result needs to be confirmed by further measurements.

Figure 4 shows the 35 meV phonon peak at nearly the same reduced wavevector measured in two different Brillouin zones. The most obvious feature is the different intensities in the two zones; this is not anomalous as it can be explained by different structure factors. However, the data also show that the linewidth at $L$ = 4.87 is significantly larger than at $L$ = 6.87. Since the linewidth is a property of the phonon that is independent of its structure factor, it implies that the 35 meV peak originates from more than one phonon. One possible explanation is that the $c$-axis motion of Fe hybridizes with the in-plane motion. Density functional theory predicts that this in-plane branch should appear around 35.5-36 meV. Thus we may see a peak due to primarily $c$-axis polarized Fe motion and another peak due to primarily $a$-axis Fe motion with some $c$-axis character mixed in. Such a mixing cannot occur if in-plane components of $Q$ are zero, but may appear if they are nonzero even if they are small such as at $Q$ = (0.04, 0.04, 4.87). The two peaks can then have different intensity variation as a function of $L$ resulting in different relative intensities at different $L$ as shown in Fig. 4. This scenario is confirmed by the fact that the phonon is well described by a single resolution-limited Lorentzian at $Q$ = (0, 0, 5.3) where $H$ = $K$ = 0 and no mixing with the in-plane modes can occur.

Figure 5 shows a comparison of our phonon results with DFT calculations of the phonon frequencies. The DFT calculations were carried out using the (GGA) combined with the linear-response technique. The calculation is in good agreement with all phonon frequencies and dispersions, except the predicted frequencies of the As-As mode are 20% higher than the measured ones throughout the Brillouin zone. Fukuda, *et al*. [15] performed similar IXS measurements on in-plane vibrations in the (1111) compounds, LaFeAsO$_{1-x}$F$_x$ ($x$ = 0, 0.1) and PrFeAsO$_{1-y}$ ($y \sim 0.1$) and compared their results with LDA calculations. They found that the experimental frequencies of intra-layer vibrations were lower than the calculated ones by 20% and that decreasing the Fe-As coupling strength by 10% in the calculation reproduced the experiment. In our experiments, the frequencies of the Fe-As vibrations along the *c*-axis are similar to the calculated ones whereas the As-As vibrations are softer than the calculated ones by 20%. One may try to explain this difference between the two datasets by the differences between the (1111) and (122) compounds. However, the results of the present calculations on double-layer compounds largely agree on Fe-As frequencies with the LDA results on single-layer compounds by Fukuda, *et al*. [15]. Thus we conjecture that the two families of compounds have very similar experimental phonon frequencies as well. Further experiments that will be performed in the near future will test this assumption.

Fukuda *et al*. [15] proposed that the simplest way to reconcile theory and their experiment is by softening the Fe-As force constant as described above. Since the frequencies of both the *c*- and *a*-polarized phonons are controlled by the same Fe-As force constants, a simple 10% softening proposed by Fukuda *et al*. may account for the difference between the theory and experiment of the *a*-axis modes, but is not consistent with a good agreement for the *c*-axis Fe-As modes. Our result suggests that the apparent softening observed by Fukuda *et al*. [15] may result from screening by intra-layer electronic degrees of freedom, as opposed to a 3-D mechanism that they proposed. It is intriguing that a similar situation exists in the copper oxide superconductor YBa$_2$Cu$_3$O$_7$: DFT gives an excellent description of *c*-polarized phonons but fails for the *a*-polarized ones. [16] Nevertheless our result combined with a recent observation of the isotope effect in pnictide superconductors strengthens the suggestion of Fukuda *et al*. [15] that unconventional electron-phonon coupling should not yet be ruled out as a superconductivity mechanism.

This research was supported in part by the National Science Foundation DMR04-0781 to the University of Tennessee, and by the Basic Energy Science Division of the Department of Energy. Use of the Advanced Photon Source was supported by the U. S. Department of Energy, Office of Science, Office of Basic Energy Sciences, under Contract No. DE-AC02-06CH11357. The construction of HERIX was partially supported by the NSF under grant no. DMR-0115852. The authors benefitted from discussions with, I. I. Mazin, E. W. Plummer, P. Dai, G. Sawatzky, Y. Uemura and S. Maekawa.

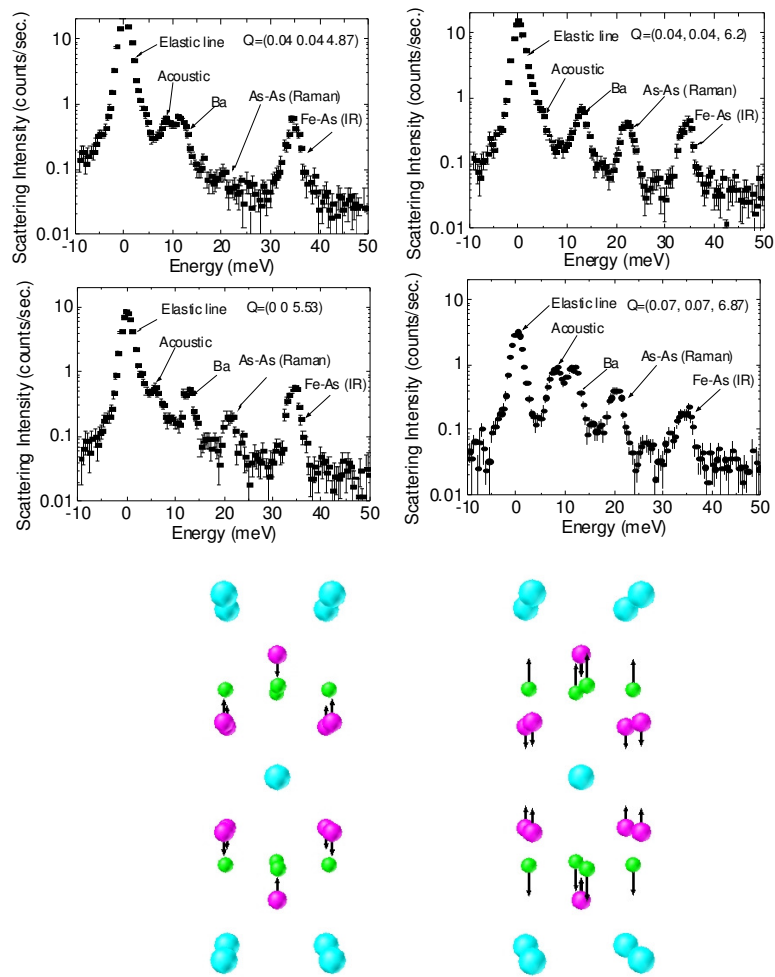

Figure 1. Top 4 panels show IXS spectra of undoped FeAs measured at four different wavevectors. The bottom row shows eigenvectors of the two highest energy zone center phonons (at 22 and 35 meV).

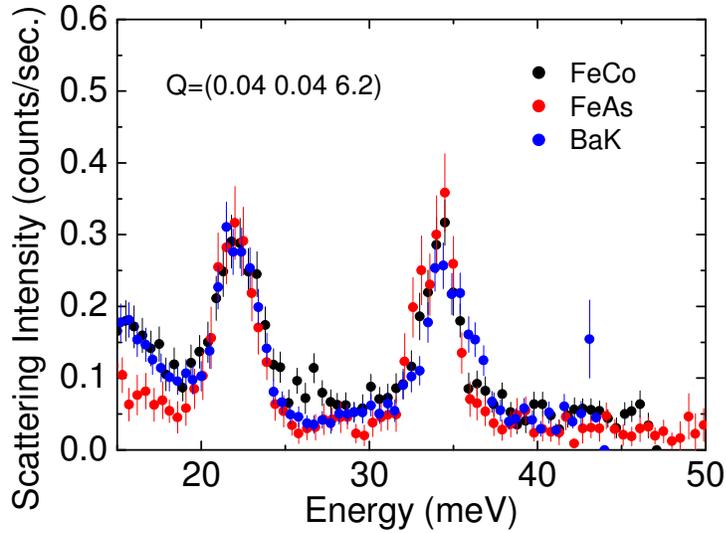

Figure 2 Composition dependence of the phonons associated with the FeAs plane. The As-As mode shows no detectable composition dependence whereas the Fe-As mode is broader and harder then the others.

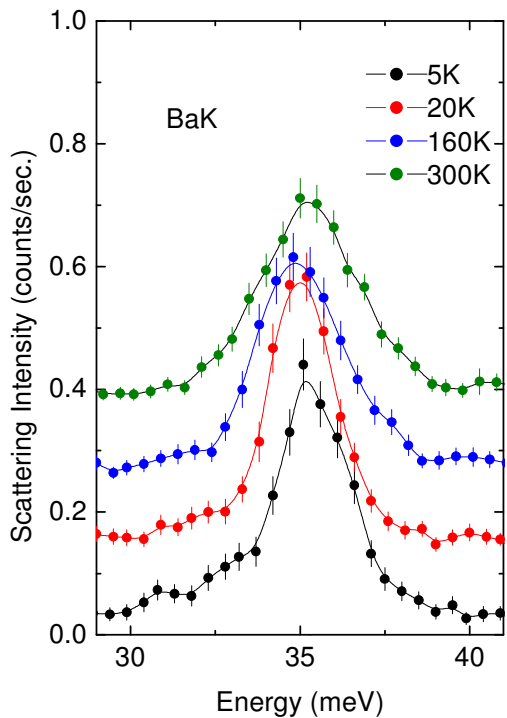

Figure 3 Temperature dependence of the Fe-As phonon at Q=(0.04 0.04 4.87) in the Ba-K sample. The mode softens by 0.5meV on cooling from 300K to 160K, it is unchanged down to 20K and then it hardens by the same amount between 20K and 5K.

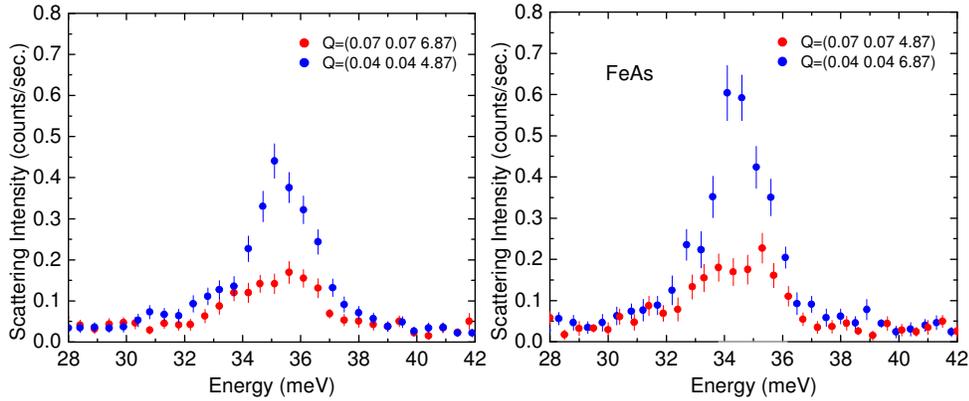

Figure 4 Unusual wavevector dependence of the Fe-As mode in the BaK and FeAs samples. The intensity at L=4.87 was too weak in the FeCo sample to make a meaningful comparison with L=6.87. The reduced reciprocal vector of the phonon is the same for the two scans, but the linewidth is different, it is broader at L=4.87. This effect is anomalous and suggests that the peak may include a contribution from two different phonons with different structure factors. This effect is present up to room temperature.

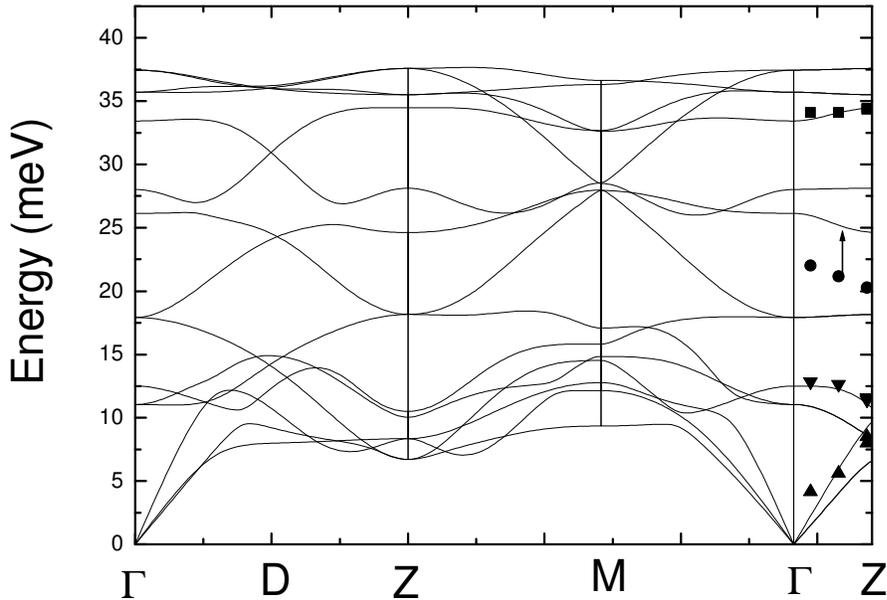

Figure 5. Comparison of the measured phonon frequencies with results of GGA. Lines represent calculated phonon dispersions, whereas the data points represent experimental phonon energies. All experimental dispersions are in an excellent agreement with the calculation, except for the branch near 20meV coming from As vibrations along the c-axis as shown in Fig. 1. Note that calculated branches that agree with the phonon frequencies have the eigenvectors consistent with the observed intensities whereas the nearby branches do not.